\documentclass[ twocolumn
]{revtex4-1}
\usepackage{comment}
\usepackage{bbold}
\usepackage{graphicx}
\usepackage{dcolumn}
\usepackage{bm}
\usepackage[version=3]{mhchem} 
\usepackage{booktabs}
\usepackage{xcolor}
\usepackage{soul}

\newcommand{\mat}[1]{\mathbf{#1} }
\newcommand{\trans}{\top}

\newcommand{\cgo}{\hat{c}}

\begin{document}

\preprint{APS/123-QED}

\title{Machine learning the DFT potential energy surface for inorganic halide perovskite CsPbBr$_3$}

\author{John C. Thomas}
\email{johnct@ucsb.edu}
\affiliation{Materials Department, University of California, Santa Barbara, Santa Barbara, CA 93106}

\author{Jonathon S. Bechtel}
\affiliation{Materials Department, University of California, Santa Barbara, Santa Barbara, CA 93106}

\author{Anirudh Raju Natarajan}
\affiliation{Materials Department, University of California, Santa Barbara, Santa Barbara, CA 93106}

\author{Anton Van der Ven}
\email{avdv@engineering.ucsb.edu}
\affiliation{Materials Department, University of California, Santa Barbara, Santa Barbara, CA 93106}

\date{\today}

\begin{abstract}
Structural phase transitions as a function of temperature dictate the structure--functionality relationships in many technologically important materials.
Harmonic Hamiltonians have proven successful in predicting the vibrational properties of many materials. 
However, they are inadequate for modeling structural phase transitions in crystals with potential energy surfaces that are either strongly anharmonic or non-convex with respect to collective atomic displacements or homogeneous strains. 
In this paper we develop a framework to express highly anharmonic first-principles potential energy surfaces as polynomials of collective cluster deformations. 
We further adapt the approach to a nonlinear extension of the cluster expansion formalism through the use of an artificial neural net model. 
The machine learning models are trained on a large database of first-principles calculations and are shown to reproduce the potential energy surface with low error.
\end{abstract}

\pacs{Valid PACS appear here}
\maketitle


\section{\label{sec:intro}Introduction}

Structural phase transitions are widespread among technologically important materials.
Statistical mechanics approaches based on the quasi-harmonic approximation are well suited to describe the finite temperature thermodynamic properties of phases that reside at a local minimum in the potential energy surface (PES) of a particular compound.
The harmonic approximation, however, breaks down for high temperature phases whose symmetry coincides with a saddle point on the PES.
These phases only emerge at elevated temperature due to anharmonic vibrational excitations.
A wide variety of high temperature phases fall in this category.
These include the bcc forms of Ti, Zr and Hf\cite{Persson2000,SouvatzisRudin2008,GrimvallPersson2012}, the high temperature cubic form of ZrO$_2$\cite{Parlinski1997,Fabris2001,Carbogno2014}, hydrides such as TiH$_2$\cite{BhattacharyaVanderVen2008} and ZrH$_2$\cite{ThomasVanderVen2013} as well as many cubic perovskite phases, including halide perovskites.\cite{BechtelVanderVen2016,YangWalsh2017,MarronierRoma2017}

While direct {\it ab initio} molecular dynamics simulations can be used to study the elevated temperature properties of anharmonically stabilized phases\cite{CockayneWoicik2018}, the computational cost of density functional theory (DFT) calculations often makes such an approach intractable.
An alternative is to rely on a model that is capable of accurately interpolating and extrapolating a limited number of DFT calculated energies within Monte Carlo or molecular dynamics simulations.

Several methods have been developed to extrapolate the first-principles PES of a compound for the purpose of studying group/subgroup structural phase transitions \cite{Vanderbilt1998,RabeWaghmare1995,WojdelGhosez2013,ZhouOzolins2014,BhattacharyaVanderVen2008,ThomasVanderVen2013,Artrith2016,Artrith2017,ThomasVanderVen2018,LiOng2018}.
In the study of group/subgourp structural transitions, the PES is typically expressed as a function of descriptors of local atomic structure. It is often convenient to formulate these descriptors as nonlinear functions of atomic displacements measured relative to the highest symmetry phase participating in the transition.
However, a challenge of this approach is to determine how the PES depends on these descriptors. In particular, the descriptors must be invariant to rigid translations and rotations of the crystal, which comprise nontrivial and highly nonlinear constraints.

Traditional approaches are based upon a Taylor expansion of the PES in terms of the Cartesian components of atomic displacement vectors.
Each term of the Taylor expansion consists of a constant, that can be treated as a chemistry dependent adjustable parameter, multiplied by a polynomial of the Cartesian components of the displacement vectors belonging to clusters of sites in the crystal.
The harmonic approximation emerges as the lowest order truncation of the Taylor expansion and consists exclusively of terms corresponding to point and pair clusters of sites.
Since polynomials of the components of displacement vectors are not invariant to rigid translations and rotations of the crystal, constraints must be imposed on the expansion coefficients, which become increasingly onerous as the order of truncation of the Taylor expansion increases.

The anharmonic cluster expansion \cite{ThomasVanderVen2013} follows a similar approach in that it expresses the PES as a sum of terms that depend on deformations of clusters of sites.
However, instead of depending directly on the Cartesian components of displacement vectors, each term corresponding to a cluster of sites is expressed as a function of collective cluster deformation coordinates that are formulated to be invariant to translations and rotations of the cluster from the outset.
The PES is then represented as a linear expansion of terms consisting of adjustable parameters multiplied by polynomials of the collective cluster deformation variables.
In the original formulation of the method, a pre-rotation step was required that relies on the computationally expensive Kabsch algorithm to determine the collective cluster deformation variables from the individual displacement vectors of the  sites belonging to each cluster.

The aims of this contribution are two fold. First we introduce new descriptors of cluster deformations that are invariant to rigid translations and rotations of clusters of sites and that can be evaluated rapidly without the need to resort to the Kabsch algorithm.
Secondly we explore the use of neural networks to identify the optimal functional dependence of the PES on the collective cluster deformation variables.
As a model system, we focus on the halide perovskite CsPbBr$_3$, a representative compound from a very promising family of chemistries for photovoltaic applications.
The perovskite form of CsPbBr$_3$ undergoes a series of group/subgroup structural phase transitions involving octahedral tilts upon cooling, adopting a cubic symmetry at high temperature, transforming upon cooling to tetragonal symmetry at 403\,K, and transforming upon further cooling to its groundstate orthorhombic phase at  361\,K\cite{RodovaNitsch2003}.
Since the different phases of inorganic halide perovskites can be connected by symmetry-lowering displacement modes from the high temperature cubic phase, it is convenient to parameterize the energy landscape in terms of distortions of the cubic reference.



\begin{figure}[t!]
 \includegraphics[width=8.0cm,keepaspectratio]{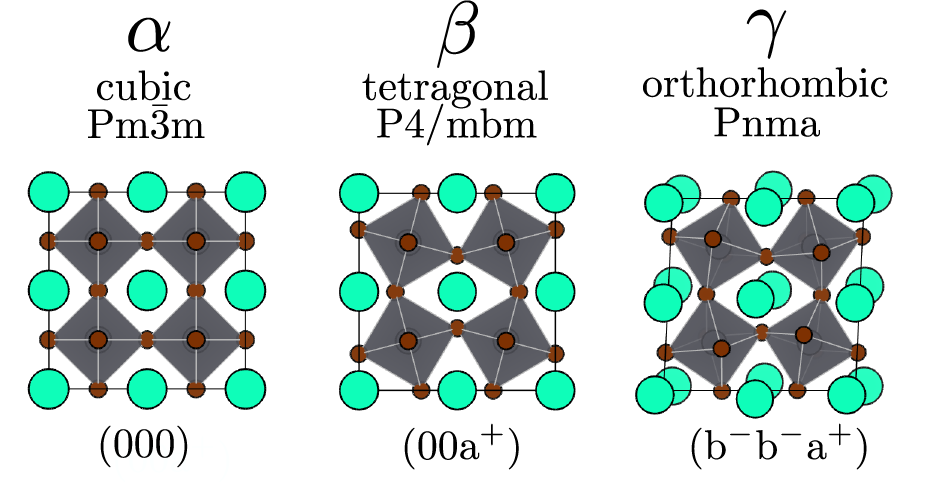}
    \caption{High temperature cubic $\alpha$-phase, intermediate temperature tetragonal $\beta$-phase, and low temperature orthorhombic $\gamma$-phase of the CsPbBr$_3$ perovskite which is the same phase sequence for many inorganic halide Cs-based perovskites.}
  \label{fgr:phases}
\end{figure}

\section{Method}

We start with the anharmonic cluster expansion approach to representing the PES of a crystal.
Within this approach, the energy of a crystal as a function of atomic displacements, $\vec{u}_i$, relative their sites, $i$, in a high symmetry reference crystal is expressed as
\begin{equation}
    E\left(\dots,\vec{u}_i,\dots\right)=E_{o}+\sum_{\alpha} \Phi^{\alpha}\left(q^{\alpha}_{1},\dots,q^{\alpha}_{N_{\alpha}}\right)
\label{eq:energy_ce}
\end{equation}
where $E_{o}$ is the energy of the reference crystal and the $\Phi^{\alpha}\left(q^{\alpha}_{1},\dots,q^{\alpha}_{N_{\alpha}}\right)$ are functions associated with clusters of sites $\alpha$.
The variables $\vec{Q}^{\alpha}=(q^{\alpha}_{1},\dots,q^{\alpha}_{N_{\alpha}})$ are functions of the displacements, $\vec{u}_i$, of the sites of the cluster $\alpha$, and  uniquely describe the degree with which cluster $\alpha$ is distorted relative to its state in the reference crystal.
The deformation variables, $\vec{Q}^{\alpha}$, must be invariant to rigid translations and rotations of the cluster to ensure that the energy of the crystal is itself invariant to rigid translations and rotations.
Both the deformation variables, $\vec{Q}^{\alpha}$, and the cluster interaction functions, $\Phi^{\alpha}$, are zero in the reference crystal.
The clusters that appear in Eq. (\ref{eq:energy_ce}) usually consist of compact, non-overlapping multi-body clusters such as tetrahedra or octahedra as well as terms for a number of longer-range pair clusters.

In the anharmonic cluster expansion of ref \cite{ThomasVanderVen2013}, the cluster interaction functions $\Phi^{\alpha}$ are expressed as an expansion of cluster basis functions according to
\begin{equation}
    \Phi^{\alpha}\left(q^{\alpha}_{1},\dots,q^{\alpha}_{N_{\alpha}}\right)=\sum_{m} V^{\alpha}_{m}\phi^{\alpha}_{m}\left(q^{\alpha}_{1},\dots,q^{\alpha}_{N_{\alpha}}\right)
\label{eq:cluster_functions}
\end{equation}
where the $\phi^{\alpha}_{m}\left(q^{\alpha}_{1},\dots,q^{\alpha}_{N_{\alpha}}\right)$ are polynomials of the elements of $\vec{Q}^{\alpha}$ and are formulated to be invariant to symmetry operations that map the reference crystal onto itself.
The expansion coefficients, $V^{\alpha}_{m}$, are determined by the chemistry of the compound and can be treated as adjustable parameters to reproduce DFT energies calculated for a sufficiently large training set of vibrational excitations relative to the reference crystal.
The requirement that the cluster basis functions, $\phi^{\alpha}_{m}\left(q^{\alpha}_{1},\dots,q^{\alpha}_{N_{\alpha}}\right)$, are invariant to symmetries of the reference crystal ensures that the energy of any two distortion fields $\left(\dots,\vec{u}_i,\dots\right)$ and $\left(\dots,\vec{u}'_i,\dots\right)$ that are related by a symmetry operation of the reference crystal have the same energy when evaluated with Eq. (\ref{eq:energy_ce}).
Polynomial basis functions, $\phi_{m}^{\alpha}$ extend to arbitrary order, but in practice only terms up to order 4 or 6 in terms of the elements of $\vec{Q}^{\alpha}$ are kept.

In the next sections, we introduce a new set of collective cluster deformation variables $\vec{Q}^{\alpha}$ that uniquely describe deformations of a cluster $\alpha$ and that are also invariant to any rigid translation or rotation of the cluster. We then introduce an approach that relies on neural networks to train cluster interaction functions $\Phi^{\alpha}\left(q^{\alpha}_{1},\dots,q^{\alpha}_{N_{\alpha}}\right)$ that go beyond a linear expansion of cluster basis functions as in Eq. (\ref{eq:cluster_functions}).

\subsection{Collective cluster deformation variables and symmetry invariant descriptors of deformation}

\subsubsection{Pair distances as measures of cluster deformations}

The starting ingredient to construct robust collective cluster deformation variables are the collection of all pair distances between the sites of a cluster $\alpha$.
This is motivated by the following property:
Given the set of all distances between pairs of atoms in a particular deformed cluster $\alpha$, it is possible to exactly reconstruct the full geometry of $\alpha$, to within a rigid rotation and translation\cite{HavelCrippen1983}.
We introduce $\vec{D}^{\alpha}=(d_{1},\dots, d_{l}, \dots, d_{N_{\alpha}})$  as comprising the pair distances $d_{l}$ between sites of a $n_{\alpha}$-point cluster where $l$ indexes unique $i$,$j$ pairs of the cluster and where $N_{\alpha}$ is the number of unique pairs in a $n_{\alpha}$-point cluster (i.e. $N_{\alpha}=n_{\alpha}(n_{\alpha}-1)/2$).



\begin{figure}
\includegraphics[width=8cm]{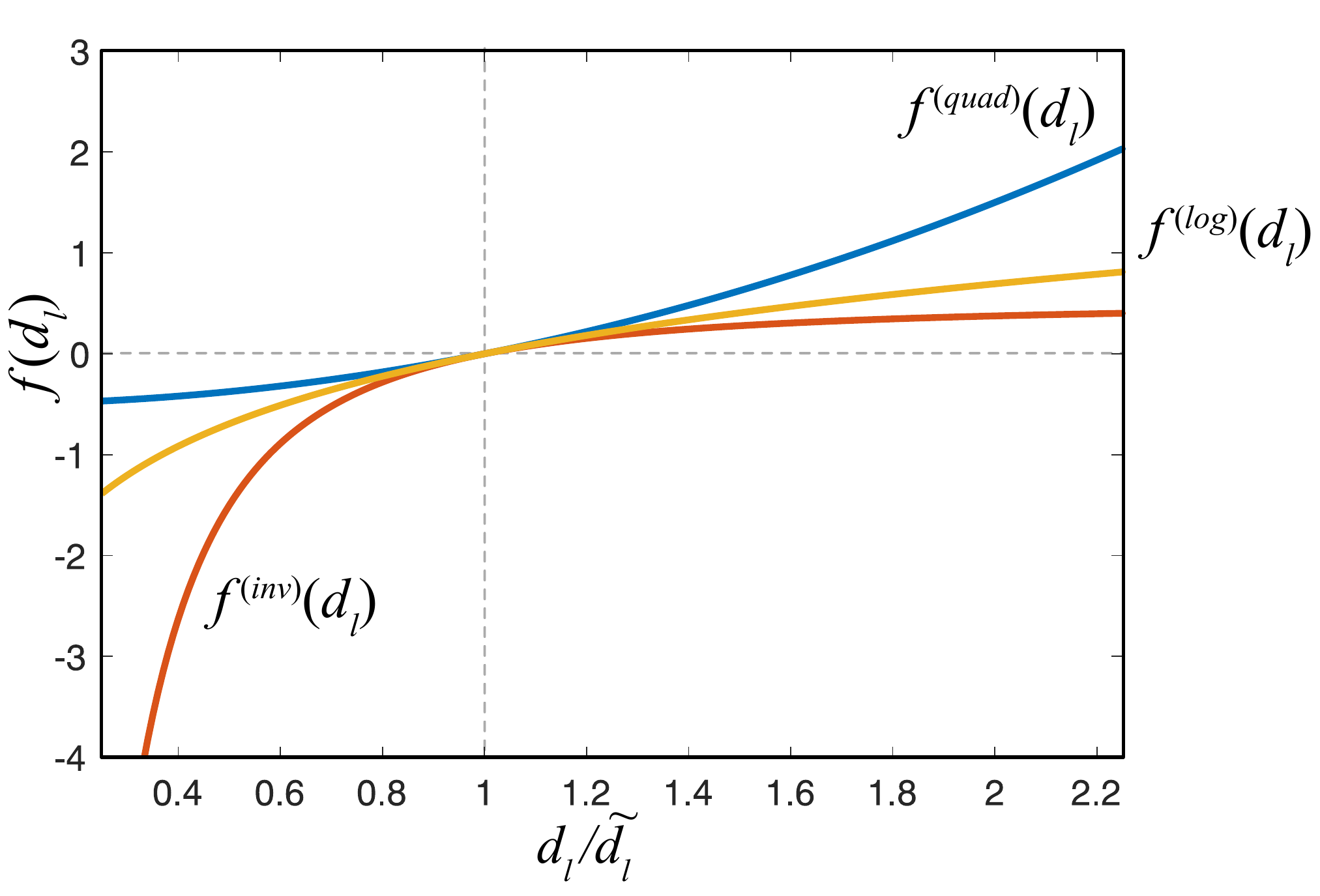}
\caption{\label{fig:DistFuncs} Illustration of difference deformation metric functions (Figure will change to match text)}
\end{figure}

We are not limited to pair distances in constructing rotationally- and translationally-invariant deformation metrics for clusters of atoms.
Any smooth monotonic function of the pair distances that can be inverted to obtain pair distances can also be used to define deformation metrics. We can thus fully specify the cluster geometry via the vector $\vec{F}^{\alpha}=(f_{1}, \dots, f_{l}, \dots, f_{N_{\alpha}})$, where $f_{l}=f(d_{l})$.
A simple choice for a deformation metric is the linear function $f^{(lin)}(d_{l})=(d_{l}/\tilde{d}_{l}-1)$, where $\tilde{d}_{l}$ is the length of the pair $l$ in the reference crystal, though other functional forms have their own advantages, such as
\begin{align}
&f^{(quad)}(d_{l}) &=& (d^2_{l}/\tilde{d}^{2}_{l}-1)/2, \\
&f^{(log)}(d_{l}) &=&\ln(d_{l}/\tilde{d}_{l}), \mathrm{\, and} \\
&f^{(inv)}(d_{l}) &=& (1-\tilde{d}^{2}_{l}/d^2_{l})/2.
\end{align}
These functions, which are depicted in Fig.~\ref{fig:DistFuncs}, all become equal to zero in the reference state (i.e. when $d_{l}=\tilde{d}_{l}$) and have identical slopes in the vicinity of the reference distance, thereby being equivalent for very small deformations. However, the behavior of each function is quite distinct at large deformations, as $d_{l}\rightarrow 0$ or $d_{l}\rightarrow \infty$.

\subsubsection{Collective cluster deformation variables}

While the vector $\vec{F}^{\alpha}$ fully determines the deformation state of the cluster, any linear transformation
\begin{equation}
\label{eqn:CDPTransform}
\vec{Q}^{\alpha}=\mat{U}\, \vec{F}^{\alpha},
\end{equation}
where $\mat{U}$ is a full-rank $N_{\alpha}\times N_{\alpha}$ matrix, yields a vector $\vec{Q}^{\alpha}$ that also fully describes the cluster deformation state.
A suitable choice for $\mat{U}$ is motivated by symmetry considerations.

The symmetries of a cluster are described by the cluster point group, which are the set of rotations and/or reflections, centered at the cluster, that map the cluster onto itself. \footnote{Physically, the atoms must be chemically identical and are thus indistinguishable, but we may assign each atom and position a distinguishing label in order to analyze the effect of symmetry.} For a cluster embedded in a crystal, the cluster point group must also leave the crystal unchanged, and so the cluster point group is a subgroup of the crystal space group. 

Application of a cluster point group operation $\cgo$ to a cluster $\alpha$ may permute the sites of the cluster.
Formally, if the reference coordinates of the cluster are columns of the $3\times n_{\alpha}$ matrix $\mat{R}^{\alpha} = (\vec{r}^{\alpha}_1 | \dots | \vec{r}^{\alpha}_{n_{\alpha}})$, then action of $\cgo$ can be expressed as
\begin{equation}
\cgo\left[\mat{R}^{\alpha} \right] = \mat{S}(\cgo)\,\mat{R}^{\alpha} + \mat{T}(\cgo)
\end{equation}
where $ \mat{S}(\cgo)$ is an orthogonal $3\times 3$ matrix (i.e., rotation, reflection, or rotoreflection) and $\mat{T}(\cgo)$ is a $3\times n_{\alpha}$  translation matrix.
The effect of a symmetry operation of the cluster on its reference coordinates is simply to permute the coordinates.
This can equivalently be represented as
\begin{equation}
\label{eqn:ClustPermSym}
\cgo\left[\mat{R}^{\alpha} \right]  = \mat{R}^{\alpha}\, \mat{W}^{\trans}(\cgo),
\end{equation}
where $\mat{W}(\cgo)$ is a $n_{\alpha} \times n_{\alpha}$ permutation matrix describing the permutation of the columns of $\mat{R}^{\alpha}$.

Just as a symmetry operation, $\cgo$, permutes the coordinates of the cluster, it also permutes the order of each distinct pair $l=(i, j)$.
The application of a symmetry operation will therefore reorder the elements of $\vec{F}^{\alpha}$.
This can expressed as
\begin{equation}
\vec{F}^{\prime\alpha}= \mat{M}^{(F)}\left(\cgo\right)\, \vec{F}^{\alpha}
\label{eqn:F_permut}
\end{equation}
where $\vec{F}^{\prime\alpha}$ and $\vec{F}^{\alpha}$ represent two deformations of the reference cluster that are related to each other by the cluster point group operation $\cgo$.
The elements of $\mat{M}^{(F)}\left(\cgo\right)$ are given by
\begin{equation}
\label{eqn:DefSymRep}
\mat{M}_{(ij),(kl)}^{(F)}\left(\cgo\right)=\mat{W}(\cgo)_{ik}\mat{W}(\cgo)_{jl}.
\end{equation}
In the above equation we have used the compound indices $(ij)$ and $(kl)$ to indicate atomic pairs after and before application of symmetry, respectively.
The symmetry representation $\mat{M}^{(F)}\left(\cgo\right)$ is also a permutation matrix that describes the discrete exchange of atomic pairs due to application of symmetry.

By combining Eq. (\ref{eqn:CDPTransform}) and Eq. (\ref{eqn:F_permut}), we can determine the effect of the application of $\cgo$ on the collective cluster deformation variables, $\vec{Q}^{\alpha}$ according to
\begin{equation}
\label{eqn:SymRepQ}
\cgo\left[ \vec{Q}^{\alpha} \right] = \mat{U}\, \mat{M}^{(F)}\left(\cgo\right)\,\mat{U}^{-1} \, \vec{Q}^{\alpha} = \mat{M}^{(Q)}\left(\cgo\right) \, \vec{Q}^{\alpha},
\end{equation}
where $\mat{M}^{(Q)}\left(\cgo\right)$ is the matrix representation describing the action of $\cgo$ on $\vec{Q}^{\alpha}$.

Equation (\ref{eqn:SymRepQ}) motivates a choice for the matrix $\mat{U}$ relating the sought after collective cluster deformation variables, $\vec{Q}^{\alpha}$, to the elements of $\vec{F}^{\alpha}$, which are each individually a function of a pair distance in the cluster.
We will use the matrix $\mat{U}$ that simultaneously block diagonalizes all the symmetry matrices $\mat{M}^{(Q)}\left(\cgo\right)$ of the cluster point group.
This choice for $\mat{U}$ generates collective cluster deformation variables $\vec{Q}^{\alpha}$ that reside in subspaces that transform under symmetry according to the irreducible representations of the cluster point group\cite{Dresselhaus2008}.
Not only does this choice simplify the formulation of polynomials of the elements of $\vec{Q}^{\alpha}$ that are invariant to the symmetry of the crystal, but it also ensures that the $\vec{Q}^{\alpha}$ can serve as order-parameters with which to detect group/subgroup symmetry breaking transitions \cite{Dresselhaus2008,ThomasVanderVen2017,ThomasVanderVen2017b}.
The elements of $\mat{U}$ for a tetrahedron cluster (assuming cluster point group $T_d$) and an octahedron cluster (assuming cluster point group $O_h$) are provided in the supporting information\cite{suppl}.

\subsubsection{Visualizing collective cluster deformations\label{sec:CDPViz}}
\begin{figure}[t!]
 \includegraphics[width=8.25cm,keepaspectratio]{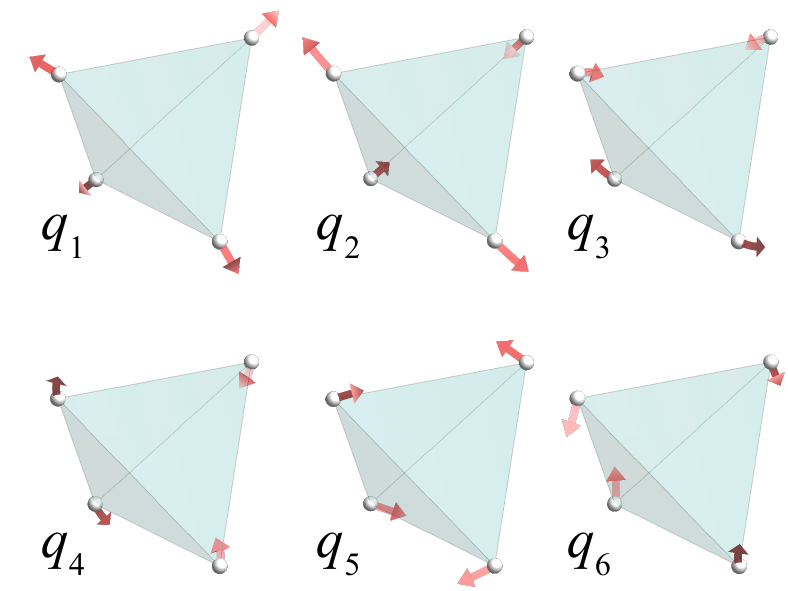}
    \caption{Visualization of the six CCDs of a tetrahedron having $T_d$ symmetry.}
  \label{fig:TetCCDs}
\end{figure}

\begin{figure}[t!]
 \includegraphics[width=8.25cm,keepaspectratio]{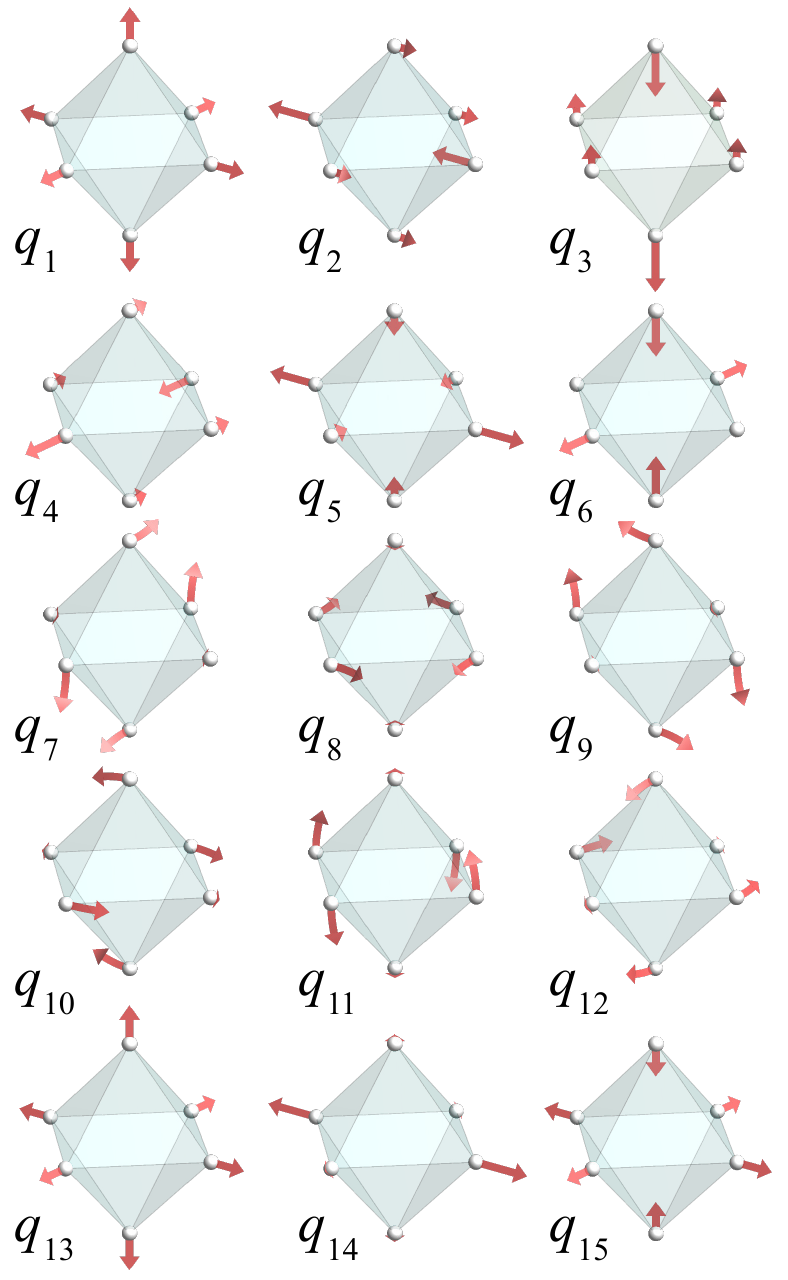}
    \caption{Visualization of the 15 CCDs of a tetrahedron having $O_h$ symmetry.}
  \label{fig:OctCCDs}
\end{figure}

We can visualize the collective distortions that are activated upon independently varying a particular CCD component $q^{\alpha}_n$ by superimposing unit vectors proportional to $\partial \vec{r}_{i}^{\alpha}/ \partial q^{\alpha}_n |_{q^{\alpha}_{m\neq n}=0}$ at each site $i$ of the cluster.
While these partial derivatives cannot be calculated directly, they can be obtained by inverting the Jacobian matrix whose elements are $J_{ij}(\vec{R}^{\alpha})=\partial{ q^{\alpha}_i}/\partial r^{\alpha}_j |_{\vec{R}^{\alpha}}$\footnote{Because the the vector $\vec{Q}^{\alpha}$ is invariant to rigid rotation and translation of the cluster, the Jacobian is rank deficient and cannot be inverted in the conventional sense. We instead utilize the Moore-Penrose inverse, which ensures that the resulting basis vectors are orthogonal to the generators of rigid translation and rigid rotations of the cluster.}.  The inverse of the Jacobian has elements $[J^{-1}(\vec{R}^{\alpha})]_{ij}=\partial{ \vec{r}^{\alpha}_i}/\partial q^{\alpha}_j |_{\vec{Q}^{\alpha}}$.

Figure \ref{fig:TetCCDs} shows the collective deformation modes corresponding to each element of $\vec{Q}^{\alpha}$ for a tetrahedron cluster. There are six such modes, with $q^{\alpha}_1$ corresponding to volumetric (i.e., symmetry-preserving) deformation. The modes corresponding to $(q^{\alpha}_2,q^{\alpha}_3,q^{\alpha}_4)$ belong to the $T_2$ irrep of $T_d$, and capture symmetry breaking to trigonal, orthorhombic, and monoclinic point groups. The modes corresponding $(q^{\alpha}_5,q^{\alpha}_6)$ belong to the $E$ irrep of $T_d$, and capture symmetry breaking to tetragonal and orthorhombic point groups.

Figure \ref{fig:OctCCDs} shows the collective deformation modes corresponding to each element of $\vec{Q}^{\alpha}$ for a six-point octahedron cluster having $O_h$ point symmetry in its reference state. There are 15 such modes, with $q^{\alpha}_1$ and $q^{\alpha}_{13}$ corresponding to volumetric (i.e., symmetry-preserving) deformation. The modes corresponding to $(q^{\alpha}_2,q^{\alpha}_3,q^{\alpha}_4)$ belong to the $T_{1u}$ irrep of $O_h$. The modes corresponding to $(q^{\alpha}_5, q^{\alpha}_6)$ and $(q^{\alpha}_{14}, q^{\alpha}_{15})$ belong to the $E_g$ irrep of $O_h$. The modes corresponding to $(q^{\alpha}_7, q^{\alpha}_8, q^{\alpha}_9)$ belong to the $T_{2g}$ irrep of $O_h$, while the modes corresponding to $(q^{\alpha}_{10}, q^{\alpha}_{11}, q^{\alpha}_{12})$ belong to the $T_{2u}$ irrep.

\subsubsection{Redundancy of cluster deformations parameters\label{sec:CDPRed}}

A non-planar cluster in three dimensions comprising $n_{\alpha}$ sites has $3n_{\alpha}-6$ deformational degrees of freedom after removal of the six rigid translational and rotational degrees of freedom.
The dimension of the CCD vector $\vec{Q}^{\alpha}$, in contrast, is $n_{\alpha}(n_{\alpha}-1)/2$.
This means that the number of CCD variables will be greater that the number of independent deformational degrees of freedom when $n_{\alpha}$ is greater 4.
The realizable values of the CCDs then reside on a $3n_{\alpha}-6$ dimensional surface (differentiable manifold) within the $n_{\alpha}(n_{\alpha}-1)/2$ dimensional space spanned by the CCDs.
This suggests a degree of redundancy among the CCD variables whereby only $3n_{\alpha}-6$ of the $n_{\alpha}(n_{\alpha}-1)/2$ CCD variables are strictly necessary to characterize the deformation state of the cluster.
While this is generally the case when the CCD variables are used to track deformations that preserve the topology of the reference cluster, there are situations where all CCD values are necessary to precisely reconstruct the geometry of the deformed cluster.

As an example, the octahedron cluster depicted in Fig.~\ref{fig:OctCCDs}  has 15 distinct CCDs but only 12 degrees of freedom. The CCDs $q^{\alpha}_1$ and $q^{\alpha}_{13}$ are qualitatively similar, as are the pairs of CCDs $(q^{\alpha}_5, q^{\alpha}_6)$ and $(q^{\alpha}_{14}, q^{\alpha}_{15})$. In addition to having identical symmetry properties, these paired sets of CCDs also describe qualitatively identical deformation modes, as demonstrated by their visualized deformation trajectories in Fig.~\ref{fig:OctCCDs}.
The nature of redundant CCDs is described in more detail in the appendix, where a procedure is outlined to identify the most important CCD variables for topology preserving deformations.

\subsubsection{Symmetry invariant polynomials of the collective cluster deformation variables}

The collective cluster deformation variables, $\vec{Q}^{\alpha}$, are constructed to be invariant to rigid translations and rotations of the cluster and have been symmetry adapted such that they transform according to the irreducible representations of the cluster point group.
The next task is to generate the polynomial basis functions, $\phi^{\alpha}_{m}(\vec{Q}^{\alpha})$, appearing in Eq. (\ref{eq:cluster_functions}).
These functions are to be invariant to the point group symmetry of the cluster.
Polynomial basis functions that are invariant to all symmetry operations of a point group that act on the arguments of the polynomial can be generated algorithmically using the Reynolds operator.
This is described in \cite{ThomasVanderVen2013, ThomasVanderVen2017}.
For the $\phi_{m}^{\alpha}$ basis functions, the approach requires the symmetry representations, $\mat{M}^{(Q)}\left(\cgo\right)$, of each cluster point group symmetry operation $\cgo$ that acts on $\vec{Q}^{\alpha}$. Symmetry-invariant CCD polynomials for the ideal 4-site tetrahedron cluster and the ideal 6-site octahedron cluster  are provided in the supporting information.

\subsection{Machine learning the potential energy landscape of a crystal}

The anharmonic cluster expansion, Eqs. (\ref{eq:energy_ce}) and (\ref{eq:cluster_functions}), can serve to interpolate and extrapolate the energies of a limited subset of first-principles calculations of different vibrational excitations of the reference crystal.
The adjustable parameters $V_{m}^{\alpha}$ that appear in Eq. (\ref{eq:cluster_functions}) can be fit to a training set of DFT energies using a variety of approaches that are commonly used to parameterize other lattice models such as alloy cluster expansions \cite{SanchezGratias1984,MuellerCeder2009,NelsonOzolins2013,VanderVenNatarajan2018}.

An alternative approach, that we pursue here, is to machine learn the PES as a function of descriptors that measure the distortion of the reference crystal.
We rely on the anharmonic cluster expansion as a starting point. Descriptors of crystal distortions must satisfy several invariance relationships. First, they must be invariant to rigid translations and rotations of the crystal. Second, they must be invariant to the space-group symmetries of the reference crystal to ensure that symmetrically equivalent deformation states of the crystal evaluate to the same energy. If the descriptors do not satisfy these constraints, they would need to be learned, necessitating a much larger training set.

The anharmonic cluster expansion can guide the identification of suitable descriptors.
While the collective cluster deformation variables $\vec{Q}^{\alpha}$ are invariant to rigid translations and rotations, they are not invariant to the symmetry operations of the crystal.
The polynomial basis functions, $\phi_{m}^{\alpha}(\vec{Q}^{\alpha})$, appearing in Eq. (\ref{eq:cluster_functions}), however, are invariant to the symmetry of the crystal and evaluate to the same value for all symmetrically equivalent cluster deformations.
Since they are a function of the CCDs, they are also invariant to rigid translations and rotations.
A sufficient number of cluster basis functions, $\phi_{m}^{\alpha}$, can therefore serve as a finger print for each symmetrically distinct distortion state of a cluster.

The approach we follow to machine learn the PES will rely on Eq. 1, but will relax the linearity of the expansion in Eq. 2.
Instead of expressing the cluster interaction functions $\Phi^{\alpha}$ as a linear expansion of the cluster basis functions, $\phi_{m}^{\alpha}$, we will train a model that has a non-linear dependence on the basis functions $\{\phi_{m}^{\alpha}\},$ which are themselves functions of the CCDs, $\vec{Q}^{\alpha}$.
We will explore two architectures for this model: a cluster-centric architecture and a site-centric architecture, relying on artificial neural networks in either case to approximate the nonlinear functional form of $\Phi^{\alpha}$.

\subsubsection{Cluster-based neural net}

To set up a cluster-based neural net description of the PES, we first rewrite Eq. (\ref{eq:energy_ce}) in a manner that exploits the symmetries of the reference crystal.
Many clusters of sites in the reference crystal are equivalent to each other by a space group operation of the reference crystal. 
For a cluster $\alpha$, we denote the set of all symmetrically equivalent clusters by $\Omega({\alpha})$, referred to as the \emph{orbit} of cluster $\alpha$.
By symmetry, all clusters belonging to a particular orbit $\Omega({\alpha})$ will have the same cluster interaction function $\Phi^{\Omega({\alpha})}$.
The anharmonic cluster expansion can then be rewritten as
\begin{equation}
    E\left(\dots,\vec{u}_i,\dots\right)=E_{o}+\sum_{\alpha}  \Phi^{\Omega{({\alpha})}}\left(q^{\alpha}_{1},\dots,q^{\alpha}_{N_{\alpha}}\right).
\label{eq:energy_ce2}
\end{equation}
Importantly, this expression indicates that although a particular cluster, such as a nearest-neighbor Pb--Pb pair, is repeated in all directions throughout the crystal, its pair interaction function can be reduced to a single functional form, $\Phi^{\Omega({\alpha})}$, that is then evaluated locally for each equivalent cluster.

Instead of relying on the linear expansion for  $\Phi^{\Omega({\alpha})}\left(q^{\alpha}_{1},\dots,q^{\alpha}_{N_{\alpha}}\right)$, we replace it with a neural net that has as inputs, not the CCDs, but rather a sufficiently large number of cluster basis functions $\{\phi_{m}^{\alpha}\}$. The energy expression can then be written as
\begin{equation}
    E\left(\dots,\vec{u}_i,\dots\right)=E_{o}+\sum_{\alpha}\mathcal{N}^{\Omega({\alpha})}\left( \dots,\phi_{m}^{\alpha},\dots\right)
\label{eq:cluster_NN}
\end{equation}
where a separate neural net, $\mathcal{N}^{\Omega({\alpha})}$, approximates the energy contribution for each distinct cluster orbit.
A visual interpretation of the computational graph for a cluster-based neural net model is depected in Figure~\ref{fgr:nn_model}(a).

\subsubsection{Site-based neural net}

An alternative approach to representing the PES is with a site-centric expression.
To this end, we define a site-centric orbit $\Omega_{i}({\alpha})$ that contains all clusters $\beta$ that are symmetrically equivalent to cluster $\alpha$ and that also contain site $i$.
The orbit  $\Omega_{i}({\alpha})$ then contains all clusters emanating from site $i$ that are symmetrically equivalent to $\alpha$.
In terms of the site-centric orbits, we can rewrite the linear anharmonic cluster expansion as
\begin{equation}
    E\left(\dots,\vec{u}_j,\dots\right)=E_{o}+\sum_{i}\sum_{\alpha}\sum_{m}\frac{1}{n_{\alpha}}V_{m}^{\alpha}\!\sum_{\beta\in\Omega_{i}({\alpha})} \phi_{m}^{\alpha}\left(\vec{Q}^{\beta}\right)
\label{eq:energy_ce3}
\end{equation}
where the sum over $\alpha$ is restricted to include only one \emph{cluster prototype} for each symmetrically distinct cluster orbit. The outer sum is over all sites $i$ in the crystal, while the innermost sum accumulates the combined contribution from all clusters that are equivalent to $\alpha$ and that include site $i$.
The factor of $1/n_{\alpha}$ corrects for over-counting due to the fact that the contribution for an individual cluster appears once for each of its constituent sites.
Equation (\ref{eq:energy_ce3}) emerges upon combining Eq. (\ref{eq:energy_ce}) and (\ref{eq:cluster_functions}) and exploiting the linearity in Eq. (\ref{eq:cluster_functions}).


Equation (\ref{eq:energy_ce3}) motivates the introduction of site-centric correlation functions defined as
\begin{equation}
	g_{\alpha,m}^{i}= \frac{1}{n_{\alpha}}\sum_{\beta \in \Omega_{i}({\alpha})} \phi_{m}^{{\alpha}}(\vec{Q}^{\beta})
\label{eq:site_correlations}
\end{equation}
The sum extends over all clusters $\beta$ that are equivalent to $\alpha$ by a crystal space group operation and that also include site $i$, ensuring that $g_{\alpha,m}^{i}$ is invariant to the subgroup of the space group that maps site $i$ onto itself.
This property guaranties that $g_{\alpha,m}^{i}$ evaluates to the same value for all distortion fields that are related to each other by a symmetry operation of the reference crystal.
A feature vector $\vec{G}^{i} =\left(g_{\alpha,1}^{i},\dots,g_{\alpha,m}^{i},\dots,g_{\alpha',1}^{i},\dots\right)$, formed by the site-centric correlation functions serves as an arbitrarily detailed descriptor of the local distortion in the vicinity of site $i$. The feature vector can be systematically improved by increasing the variety and cutoff range of symmetrically distinct clusters, $\alpha$, $\alpha'$ etc., constituting the the descriptor, as well as the order of their corresponding basis functions.

\begin{figure}[t!]
 \includegraphics[width=8.25cm,keepaspectratio]{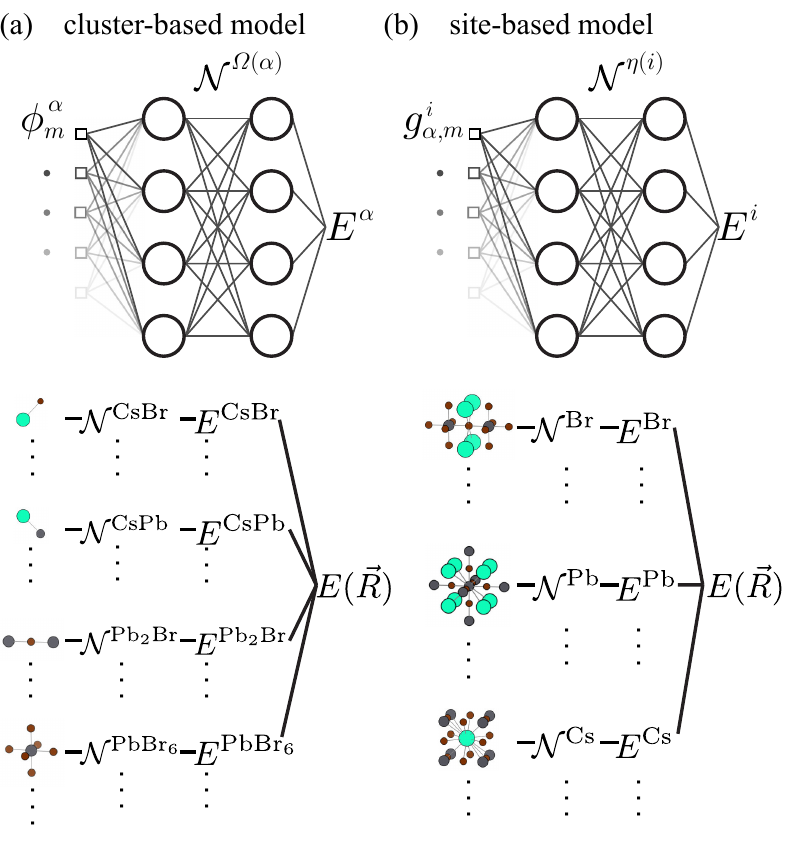}
    \caption{Visualization of how (a) site-based and (b) cluster-based models incorporate site-averaged basis functions or cluster-based basis functions respectively.}
  \label{fgr:nn_model}
\end{figure}

A site-centric neural net description of the PES is thus formulated in terms of the feature vector $\vec{G}^{i}$ according to
\begin{equation}
	 E\left(\dots,\vec{u}_j,\dots\right)=E_{o}+\sum_{i} \mathcal{N}^{\eta(i)}(\vec{G}^{i})
\label{eq:site_NN}
\end{equation}
where the total energy of the crystal is a sum over contributions from each individual site $i$. $\eta(i)$ refers to the orbit of all sites equivalent to site $i$ with respect to the symmetry of the reference crystal, such that there is a separate approximation function, $\mathcal{N}^{\eta(i)}$, for each symmetrically distinct site of the reference crystal. The site-based neural net model is summarized in Figure~\ref{fgr:nn_model}(b).

\subsubsection{Artificial Neural Network}
Whether working in the site-based or cluster-based cluster expansion, we make use of artificial neural network models that take as inputs $\vec{x}$ (where $x_i$ could be either the local-orbit summed basis functions in the site-based model, or simply the evaluated basis functions in the cluster based model)  and output an energy $e$. Artificial neural networks are hierarchical recursive functions made up of activation nodes $f_{i}$ which represents a non-linear function $f$ at node $i$. A one-layer neural net produces output $e$ from inputs $\mathbf{x}$ as follows:
\begin{align}
e = b^{(1)} +  \sum_{j}w^{(1)}_j f_j( b^{(0)}_{j} + \sum_{k} {x}_{k} {W}^{(0)}_{kj}  )
\end{align}
where $b^{(1)}$ is a bias term associated with the 1st layer, and $b^{(0)}_{j}$, are bias terms associated with the input layer into node $j$ of the first hidden layer. ${W}^{(0)}_{kj}$ represents the weight matrix connecting the input layer to the first hidden layer, and $w^{(1)}_j$ is the weight matrix connecting the hidden layer to the output layer. The model variables are the weights and biases which are trained through optimization techniques described below. The activation function, $f_{j}$,  can take several forms including the hyperbolic tangent, rectified linear unit, or logistic function. In this study we used the hyperbolic tangent exclusively.

\subsubsection{Objective Function}

In order to train the neural network model, we must minimize a convex objective function. Here we choose an objective function that penalizes the sum of the squares of the differences in model energies and those calculated with DFT for a large number of different vibrational excitations.
\begin{align}
\Gamma&= \sum_{\sigma} \left( E_{\text{ANN}}(\sigma) - E_{\text{DFT}} (\sigma) \right) ^2
\label{eq:rmse_error}
\end{align}
where $\sigma$ denotes different vibrational excitations.
The objective function is minimized with respect to the weights of the neural network. Many optimization algorithms exist to optimize the weights of the network function. We employed the Adam algorithm in this study\cite{Goodfellow2016,KingmaBa2014}.


\begin{figure}[!t]
 \includegraphics[width=8.25cm,keepaspectratio]{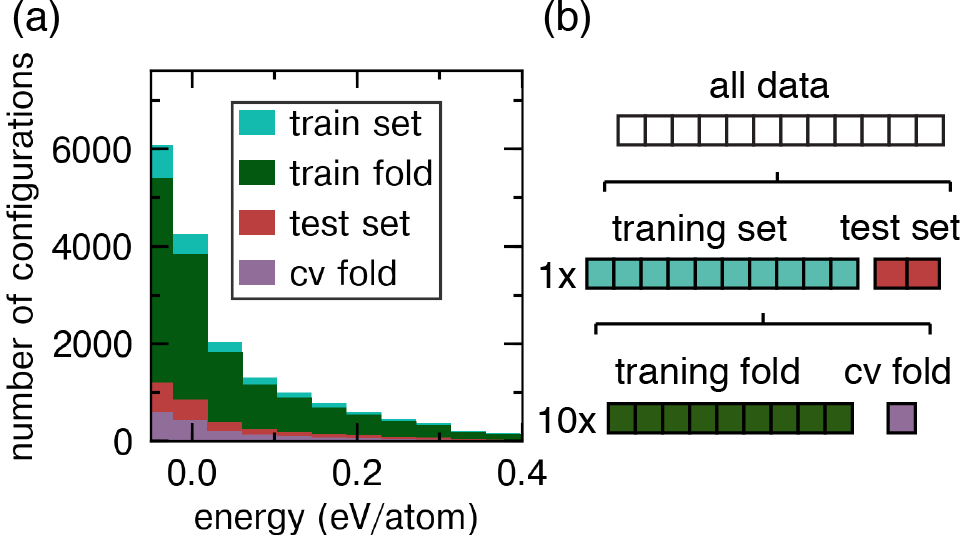}
    \caption{(a) Distribution of energies for all configurations in the database. (b) All data is split into a training set and test set. The training set is further subdivided into 10 training folds and 10 validation folds for use in hyperparameter tuning.}
  \label{fgr:energy_distribution}
\end{figure}

\section{Potential energy surface of halide perovskites}

In this section we develop a neural network model of the potential energy surface of CsPbBr$_3$, a compound belonging to a class of promising perovskite based materials for electronic and photovoltaic applications.
CsPbBr$_3$ undergoes a series of group/subgroup structural phase transitions upon cooling.
At high temperature, CsPbBr$_3$ is stable in a cubic perovskite crystal structure, but transitions to tetragonal and orthorhombic symmetries at lower temperatures due to tilting of the PbBr$_3$ octahedra.
As with many halide perovskites, the cubic and tetragonal forms of perovskite CsPbBr$_3$ correspond to saddle points on the potential energy surface of the compound \cite{Bechtel2018}.
These phase only emerge at finite temperature due to large scale anharmonic vibrational excitations.

\subsection{DFT}
Density functional theory calculations were performed using the Vienna Ab Initio Simulation Package (VASP).~\cite{KresseFurthmuller1996,Kresse1999} A plane wave basis set with an energy cutoff of 400 eV was employed and projector augmented wave psuedopotentials (PAW).~\cite{Blochl1994,KresseFurthmuller1996} The GGA-PBEsol functional was used to approximate electron correlation and exchange.~\cite{Perdew2008} Energies were converged to within 1 meV / atom with respect to k-point density and a 6$\times$6$\times$6 $\Gamma$-centered k-point mesh was used for the CsPbBr$_3$ unit cell. The VESTA program suite was used to visualize crystal structures.

\subsection{Training Set}
\begin{figure*}[t!]
 \includegraphics[width=16.25cm,keepaspectratio]{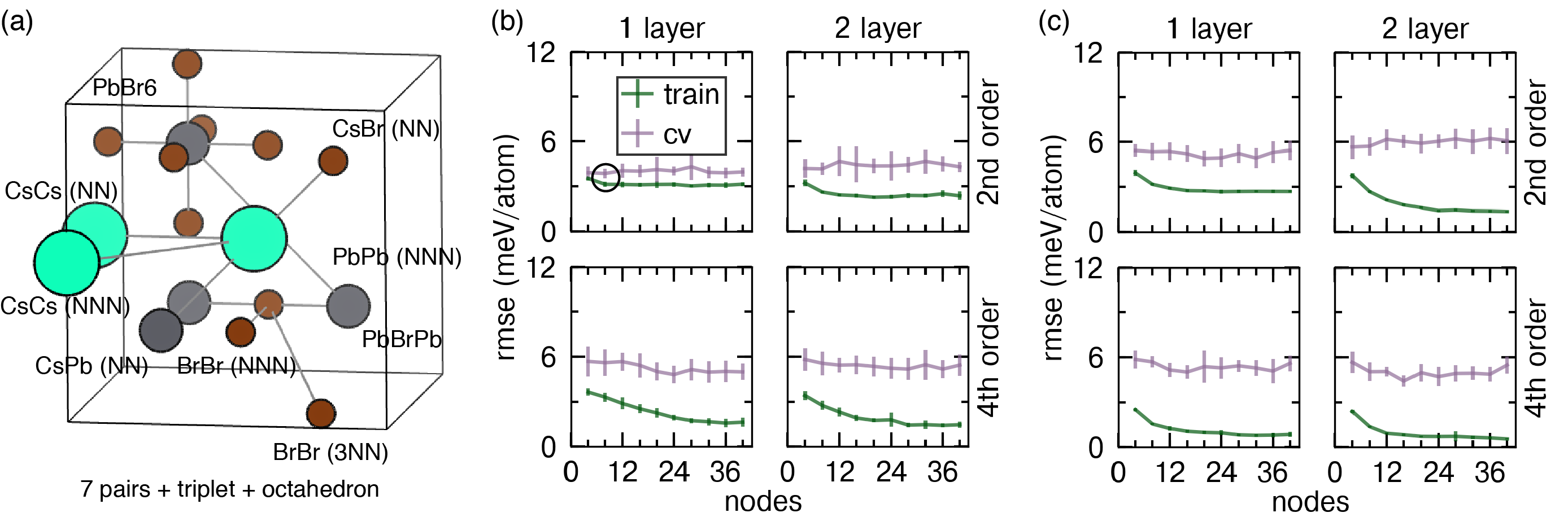}
    \caption{(a) Clusters used in final model which includes 7 pairs, 1 triplet, and 1 octahedron. Results of 10-fold cross validation for (b) cluster-based model and (c) site-based model. Training and validation (cv) average RMSE is plotted with error bars of 1 standard deviation. In (b,c), left columns indicate 1 hidden layer while right columns indicate 2 hidden layers and top rows indicate 2nd order models while the bottom row indicates 4th order models.}
  \label{fgr:cv}
\end{figure*}

The training set is a critical component in a machine learning problem.
The resulting model is only as good as the training set.
Each element of the training set, corresponding to the energy of a particular state of strain and a particular set of atomic displacements relative to the high symmetry reference state, will be referred to as a configuration, $\sigma$.
The most important regions of the PES include the potential energy wells in which the ground state structure resides. Therefore, much effort was made to sample configurations near the ground state structure along with the structures associated with the intermediate tetragonal phase and the high temperature cubic phase.

Sampling the PES was done in several ways. The starting point began with the geometric relaxation of the 15 tilt systems as previously described in \cite{Bechtel2018}. For each of these relaxed structures, systematic displacement enumerations were made in terms of symmetry-adapted displacement modes, i.e. the displacement fields that block diagonalize the crystal symmetry representation. The same supercell (2$\times$2$\times$2) was used for all of the tilt systems to avoid numerical errors incurred when using differing k-point grids. Systematic strain enumerations were also included on the primitive perovskite structure, and the irreducible wedge of each subspace was sampled in the volume 1 cell. In addition to systematic enumerations, stochastic sampling of strains and displacements were made to generate more configurations. The strain and displacement fields were chosen at random from an n-sphere, and the correlations were compared to existing configurations to ensure uniqueness, i.e. that a very similar structure wasn't already included in the database. Also interpolations between structures were used for example between the three experimentally observed phases. In total, 31,000 configurations were calculated.


\subsection{Hyperparameter tuning and model training}
Training high-quality neural network models requires the selection of optimal model hyperparameters specifying the network architecture (i.e., number and connectivity of nodes) and number of input descriptors. 
 We used $k$-fold cross validation to determine which set of hyperparameters best generalize to holdout sets of model validation data. 
 This process is similar to model selection approaches in alloy cluster expansions, where a set of cluster basis functions are chosen to minimize a cross-validation metric.
Due to the large number of input descriptors in an anharmonic cluster expansion, we restrict ourselves to five sets of clusters: (1) 4 pairs + 1 octahedron, (2) 5 pairs + 1 octahedron, (3) 8 pairs + 1 octahedron, (4) 4 pairs + 1 triplet + 1 octahedron, (5) 7 pairs + 1 triplet + 1 octahedron. 
The covalent bonding within the octahedra of CsPbBr$_3$ motivated the inclusion of an octahedral cluster. 
For each cluster in the model, we tested two groups of cluster basis functions to serve as input features: one included all cluster basis functions of the CCDs up to 2$^{nd}$ order and another included all basis functions up to 4$^{th}$ order. 
Additionally, we tested several network architectures by varying the number of hidden layers and the number of nodes per hidden layer, resulting in 400 unique hyperparameter sets. 

Given a hyperparameter set, training the weights and biases in a neural net requires an optimization scheme\cite{Goodfellow2016}. We employed a batch training strategy with batch sizes of 2, 10, 100 and 1000 with at least 1000 training epochs per batch size. The Adam optimizer was used to update model weights and biases such that the least squares error of Eq. \ref{eq:rmse_error} was minimized.\par
Validation and training sets were used to find ANN hyperparameters that resulted in the most generalizable models with the smallest error. The total data set was split into a training set (80\% of data), and a test set (20\% of data). The test set was kept removed from any training iterations such that it remained an unbiased evaluator of model performance. K-fold cross validation with 10 folds was used to find the optimal hyperparameters (number of nodes, layers, and input features in the ANN model). A model was trained on 90\% of the training dataset and a cross-validation error (CV) was evaluated on the remaining 10\%.
 This procedure is repeated 10 times leaving a different fold of the training dataset out each time.

\begin{figure*}[t!]
 \includegraphics[width=16.25cm,keepaspectratio]{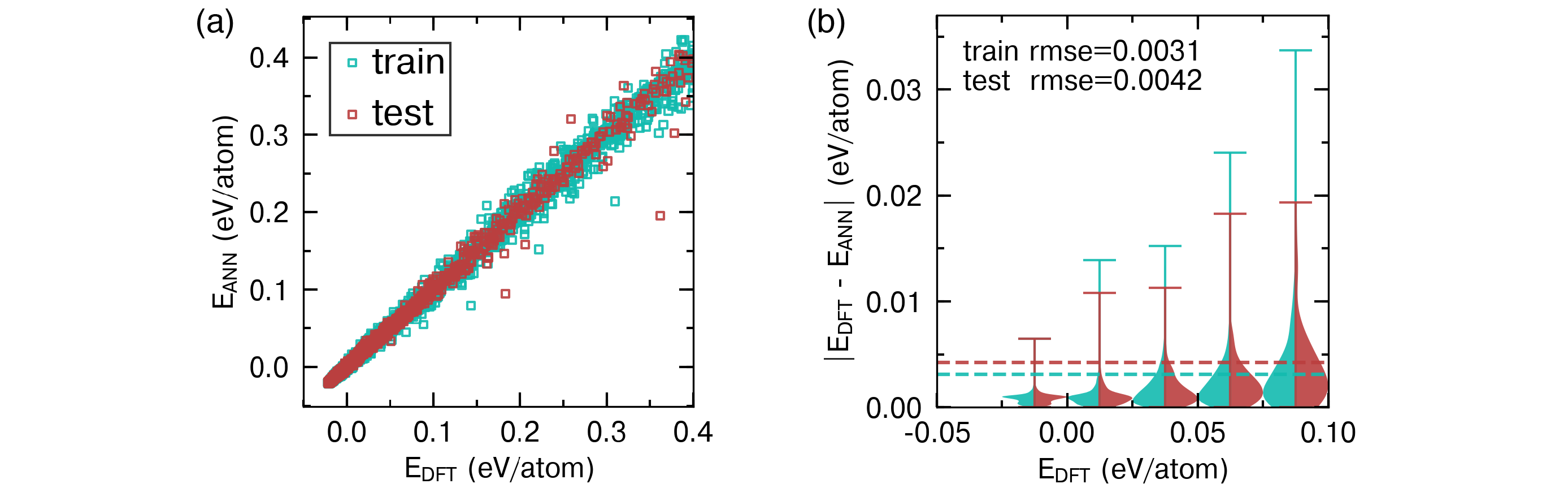}
    \caption{Fitting statistics for 1 layer 2nd order cluster-based model with 8 hidden nodes per layer. (a) ANN energy vs DFT energy shows that both training and test set show similar average error. (b) Distribution of errors for lowest 125\,meV configurations binned into 25\,meV bins. The red and green dashed lines indicate the RMSE over the entire test and training set resepectively.  Low energy configurations show very low error.}
  \label{fgr:fit_stats}
\end{figure*}
\subsection{Optimal Hyperparameters}

Figure~\ref{fgr:cv} displays the training results for the best performing set of hyperparameters. This set consists of basis functions generated from 7 pairs, 1 triplet, and 1 octahedral cluster as pictured in Figure~\ref{fgr:cv}(a).
Four other combinations of clusters were tested, but it was found that including more clusters, and especially including the triplet cluster resulted in more robust models. The neural net training results are displayed in Figures~\ref{fgr:cv} for the cluster-based model (Figures~\ref{fgr:cv} (b)) and the site-based model (Figures~\ref{fgr:cv} (c)). For each model, input features up to order 2 or order 4 basis functions were tested [rows of Figures~\ref{fgr:cv}(b,c)] as well as number of hidden layers [columns of Figures~\ref{fgr:cv}(b,c)].

The site-based and cluster-based models perform similarly with several key differences.
First, the cluster-based models generalize better to the validation set with smaller validation errors among all tested models.
However, the site-based models achieve smaller errors on the training folds.
Large differences between the training error and the validation error indicate that the models tend to overfit the training data and generalize poorly.
The order 2 cluster-based model with 1 hidden layer performed the best in terms of generalizability with both the smallest validation error and the smallest difference between the training and validation errors. In particular the model with order 2 cluster-based model with 1 hidden layer and 8 nodes per hidden layer had the smallest validation error among all models and was therefore chosen as the best model according to the cross validation scheme.

\subsection{ANN Fit Evaluation}
After finding the optimal hyperparameters for our model (shown with the black circle on Figure~\ref{fgr:cv}(b) indicating the order 2 cluster-based model with 1 hidden layer of 8 nodes), we retrained the model on the full training set and calculated the error on the holdout set as shown in Figures~\ref{fgr:fit_stats}(a,b). The training and test rmse were similar to those found in the hyperparameter tuning as expected. Additionally, we investigated the distribution of errors for different energy regions as shown in  Figures~\ref{fgr:fit_stats}(b). Interestingly, the model performs best for the lowest energy configurations, meaning that it faithfully reproduces the important ground state structures.


\begin{figure}[h!]
 \includegraphics[width=8cm,keepaspectratio]{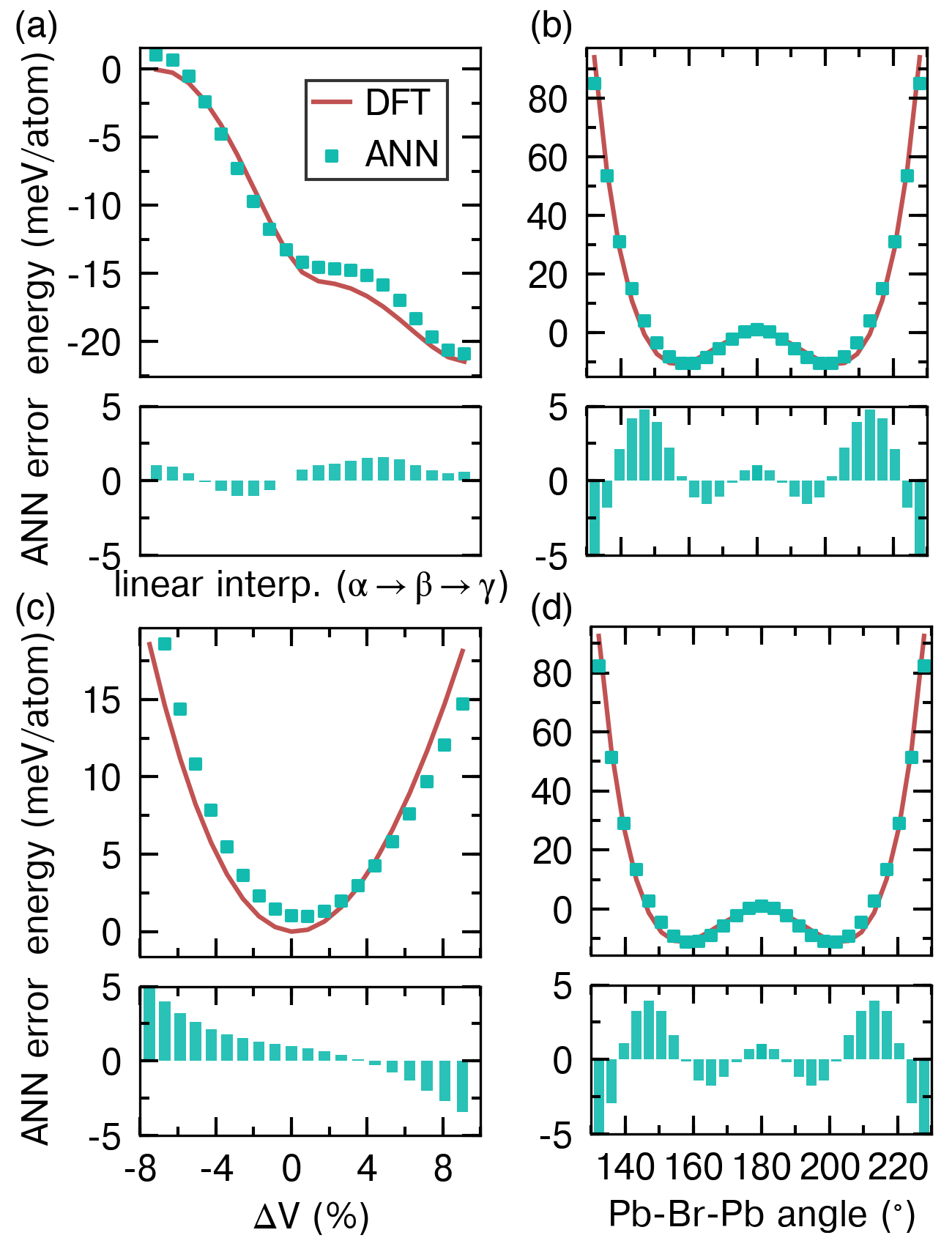}
  \caption{Model and DFT energies as a function of (a) a linear interpolation experimentally observed phases, (b) in-phase tilts, (c) volume, and (d) anti-phase tilts. In all cases, the ANN PES aligns well with the DFT energy surface.}
  \label{fgr:test_summary}
\end{figure}

Figure~\ref{fgr:test_summary} shows how the model reproduces the DFT potential energy surface along important paths in the space of atomic deformations.
Figure~\ref{fgr:test_summary} (a) shows the energy as a function of a linear interpolation between the cubic ($\alpha$), tetragonal ($\beta$) and orthorhombic ($\gamma$) phases of perovskite CsPbBr$_3$.
Also shown is the energy of the crystal as a function of (b) in-phase octahedral tilt-mode amplitude, applied to the ideal cubic structure, (c) volumetric strain deformations of the crystal lattice, and (d) anti-phase octahedral tilt amplitude, applied to the ideal cubic structure.
In all cases, the model predictions align well with the DFT energy surface.
Furthermore, the model PES tends to be relatively smooth.
The results for this simple model indicate that neural networks can reliably reproduce the potential energy surface of complex compounds such as CsPbBr$_3$, especially in the region of low energy configurations.

\section{Discussion}
A large variety of compounds adopt phases at high temperature that have symmetries coinciding with a saddle point on a zero Kelvin potential energy surface (PES).
Crystal symmetries corresponding to a saddle point of the PES are dynamically unstable at low temperature but can become stable at high temperature through large anharmonic vibrational excitations.
Phonon theories based on the harmonic approximation are unable to describe the high temperature thermodynamic properties of anharmonically stabilized phases.
Instead, Monte Carlo or molecular dynamics simulations must be used to numerically perform thermodynamic averages over vibrational microstates sampled at high temperature.
Direct simulation approaches, however, require a model of the energy of the crystal as a function of the atomic displacement and lattice strain degrees of freedom.

Mapping the atomic coordinates of a solid to reproduce a first-principles energy landscape is a challenging, high-dimensional supervised learning problem that requires careful consideration of many aspects of the machine learning pipeline, including feature engineering, model training, and model selection.
High quality input features are an essential ingredient of any machine-learned model.
In this study we have introduced collective cluster deformation (CCD) variables that uniquely describe the deformation of a particular cluster within the crystal relative to its geometry in a high symmetry reference state. 
Because the CCDs are symmetry-adapted functions of the set of all pair distances within the cluster, any model that is a function of these variables is inherently invariant to rigid-body rotation or translation of the crystal.  
Although they are defined relative to an undeformed high-symmetry crystal, the CCDs are not themselves invariant to the symmetry of this reference crystal.
As such, the feature vector forming the input layer of the neural net is constructed from symmetry-invariant polynomials of the CCD variables, thus ensuring that the learned model is invariant to these additional crystal symmetries.
Taken together, these properties specify features that are particularly well suited to machine-learning PES models for compounds that undergo group/subgroup structural transformations, in which the high symmetry phase is often stabilized at high temperature by large, anharmonic vibrational excitations.
Moreover, the CCDs are themselves useful descriptors of local structure that have potential applications in high-throughput crystallographic data-mining frameworks, such as recently described workflows for characterizing local coordination environments\cite{WaroquiersHautier2017}.

The model selection methodology described here showed that low order basis functions tended to result in more generalizable models, with low error on holdout test sets.
Higher order descriptor functions, as well as deeper (more layers) and wider (more nodes) neural nets, tended to overfit the training data resulting in poorly generalized models.
The low training error of the more complex models indicates that the descriptors provide adequate information for models to learn the DFT energy surface, and motivate further studies focusing on reducing overfitting using techniques such as dropout or weight decay.

Two approaches were introduced in this study, one based on a \textit{cluster}-centric neural net architecture and the other based on \textit{site}-centric architecture.
The cluster-based models are direct generalizations of previous anharmonic vibrational cluster expansion models as introduced by Thomas and Van der Ven.~\cite{ThomasVanderVen2013}
The current work extends the linear models in ~\cite{ThomasVanderVen2013} by allowing the functional form of the cluster energy to be learned by the machine learning model.
The site based model using cluster basis functions is an extension to vibrational energy of the site-based neural-net approach introduced by Natarajan and Van der Ven for modeling configurational energy\cite{Natarajan2018}.
One benefit of the site-based model is that it allows interaction terms between basis functions from different clusters, which may explain the lower error achieved by the site-based model.
The site-based model has some similarities to other descriptor-based machine learning approaches where descriptors are written in terms of exponentials of pair distances and bond angles\cite{Behler2007,Bartok2010}.
A key difference of the approach introduced here, however, is that it is specifically designed to represent the energy of a crystal relative to a high symmetry reference crystal, making it especially suited for studies of group/subgroup structural transitions and the thermodynamics of anharmonically stabilized phases.
The reliance on descriptors that are invariant to the symmetries of the high symmetry reference phase ensures that symmetries are automatically satisfied. However, relative to more generic approaches, CCD descriptors that are measured relative to a high-symmetry reference crystal have much higher bias (in an information-theoretical context), so that significantly more information about the crystal deformation state can be encoded by fewer descriptors.

\section{Conclusions}

The development of anharmonic vibrational hamiltonians is a challenging problem, however, by making use of machine learning techniques it is possible to capture a high degree of complexity that is present in the DFT energy landscape. We have presented a framework that utilizes neural-network models to reproduce the DFT energy landscape with high accuracy in the vicinity of a high-symmetry reference crystal. To construct features for the neural-network model we introduced collective-cluster-deformation variables, which are descriptive and easy-to-calculate functions of local geometry that are invariant to rigid-body transformations. The use of machine learning models is appealing because it removes much of the manual selection of terms in a Hamiltonian. Instead, the functional forms are learned through the training process. However, machine learning models, especially non-linear neural networks have a tendency to overfit the training data, and, therefore, hyperparameter tuning must be carefully considered. The next step in the progression of machine learning Hamiltonians is their use in finite temperature thermodynamics simulations which is a natural extension of the work presented here.

\section{Acknowledgement}
This material is based upon work supported by the National Science Foundation, Grant No. OAC-1642433. Computational resources provided by the National Energy Research Scientific Computing Center (NERSC), supported by the Office of Science and U.S. Department of Energy, under Contract DE-AC02-05CH11231, are gratefully acknowledged in addition to support from the Center for Scientific Computing from the CNSI, MRL: an NSF MRSEC (DMR-1720256).

\appendix

\section{Identifying deformation coordinates for dimensionality reduction \label{sec:GramMat}}

A $n_{\alpha}$-atom non-planar cluster has $3n_{\alpha}-6$ deformational degrees of freedom in three dimensions (after removal of rigid translation and rotation) and has $N_{\alpha}=n_{\alpha}(n_{\alpha}-1)/2$ pair distances. For $n_{\alpha}>4$, the number of pair distances exceeds the number of cluster degrees of freedom, such that any realizable deformation vector, $\vec{F}^{\alpha}$, must be confined to a $3n_{\alpha}-6$-dimensional surface. In the vicinity of the undeformed cluster, coordinates on this cluster deformation surface can be projected uniquely into a $3n_{\alpha}-6$-dimensional subspace, and a point in the subspace can be described by a truncated $3n_{\alpha}-6$-element vector of optimized CCDs, which we denote $\vec{Q}^{\star\alpha}$.

A simple linear approximation of the cluster deformation surface can be computed from the matrix image of the Jacobian $\mat{J}_{\vec{F}}(\vec{R}^{\alpha})$. However, a more robust set of linearized coordinates can be obtained by accounting for the fact that the pairwise deformation metrics are correlated for small deformations of the cluster. We define a correlation matrix $\mat{G}^{(F)}$ whose elements are the overlap, or similarity, between the deformation metrics of two pairs within the cluster. The elements of $\mat{G}^{(F)}$ are computed as inner products over the space of functions of the cluster coordinates, such that
\begin{equation}
\label{eqn:InnerProd}
G^{(F)}_{m,n}=\left< f_m, f_n \right> = \int_{\vec{R}^{\alpha}} \mathrm{d}^{3N_{\alpha}} \vec{R}^{\alpha} \, \left[ p(\vec{R}^{\alpha}) f_{m}\left(\vec{R}^{\alpha}\right) f_{n}\left(\vec{R}^{\alpha}\right) \right],
\end{equation}
where $p(\vec{R}^{\alpha})$ is a probability density over all possible geometries of cluster $\alpha$, and the integral is taken over the entire configuration space of $\vec{R}^{\alpha}$. A simple choice of $p(\vec{R}^{\alpha})$ is a $3n_{\alpha}$-dimensional multivariate normal distribution, centered at the coordinates of the undeformed reference cluster and having an isotropic variance $\sigma^2$. This definition allows an analytic expression for Eq.~(\ref{eqn:InnerProd}) for many choices of deformation metric. Physically motivated choices of the standard deviation $\sigma$ are in the range of 10-25\% of the nearest-neighbor pair distance for the crystal under consideration.

The correlation matrix $\mat{G}^{(F)}$ can be used to identify an optimized coordinate transformation from $\vec{F}^{\alpha}$ to $\vec{Q}^{\star\alpha}$.  For a given change of basis $\vec{Q}^{\alpha}=\mat{U} \,\vec{F}^{\alpha}$, the corresponding transformation that takes $\mat{G}^{(F)}$ to $\mat{G}^{(Q)}$ is
\begin{equation}
\mat{G}^{(Q)} = \mat{U}^{-\trans}\,\mat{G}^{(F)}\, \mat{U}^{-1}.
\end{equation}
The elements of $\mat{G}^{(Q)}$ measure the correlation between individual components of $\vec{Q}^{\alpha}$ over $p(\vec{R}^{\alpha})$. If the transformation $\mat{U}$ is chosen appropriately, the correlation matrix $\mat{G}^{(Q)}$ will be the identity matrix, meaning that individual CCDs have unit variance and are uncorrelated over $p(\vec{R}^{\alpha})$. This occurs when
\begin{equation}
\label{eqn:CCDTransform}
\mat{U}={\mat{G}^{(F)}}^{1/2} = \mat{\Lambda}^{1/2}\,\mat{V}^{\trans},
\end{equation}
where $\mat{\Lambda}$ is a diagonal matrix of the eigenvalues of $\mat{G}^{(F)}$ and $\mat{V}$ is an orthogonal matrix of the eigenvectors of $\mat{G}^{(F)}$. Each row $i$ of $\mat{U}$ corresponds to a linear combination of the components of $\vec{F}^{\alpha}$ that yield a particular CCD value. The $\mat{U}$ transformation matrices are provided in the supplemental information for both a 4-point tetrahedron and  6-point octahedron cluster having maximal symmetry\cite{suppl}.

The transformation matrix $\mat{U}$, as defined in Eq.~(\ref{eqn:CCDTransform}) is optimal in several ways. First, the original correlation matrix $\mat{G}^{(F)}$ is invariant to symmetry in the sense that
\begin{equation}
\hat{c}\left[\mat{G}^{(F)}\right]=\mat{M}^{(F)}(\hat{c})\,\mat{G}^{(F)} \,\mat{M}^{(F)\trans}(\hat{c})=\mat{G}^{(F)},
\end{equation}
where $\hat{c}$ is an operation in the point group of cluster $\alpha$. This invariance relation is due to the fact that symmetrically-equivalent pairs of pair-deformation metrics have identical correlation. It is well known that if a symmetric matrix, such as $\mat{G}^{(F)}$, is invariant to a group representation (e.g, $\mat{M}^{(F)}(\hat{c})$), then any transformation that diagonalizes $\mat{G}^{(F)}$ also block-diagonalizes the symmetry matrices $\mat{M}^{(F)}(\hat{c})$. This means that the resulting CCD vector space, $Q$, is naturally separable into invariant subspaces, and that, under ideal circumstances, the invariant subspaces of $Q$ will correspond to irreducible representations. To this end, eigenvectors of $\mat{G}^{(F)}$ having the same eigenvalue correspond to CCDs that are within the same invariant subspace. Moreover, if the eigenvalues of $\mat{G}^{(F)}$ are ordered from largest to smallest, there are spectral gaps between irreducible subspaces, with the largest gap occurring between the first $3n_{\alpha}-6$ eigenvalues of $\mat{G}^{(F)}$ and the remaining eigenvalues.  This gap occurs due to an underlying difference in behavior between directions with respect to the cluster deformation surface in the vicinity of the reference cluster. The first $3n_{\alpha}-6$ directions approximately follow the cluster deformation surface, and so they have much larger variance than the remaining directions, which are nearly orthogonal to the cluster deformation surface. The truncated CCD vector, $\vec{Q}^{\star\alpha}$, then corresponds to the first $3n_{\alpha}-6$ elements of the CCD vector obtained from the transformation matrix in Eq.~\ref{eqn:CCDTransform}.


\bibstyle{apsrev4-1}
\end{document}